\documentclass[12pt]{iopart}
\usepackage{graphicx} % ÒýÈëgraphicxºê°ü

\begin{document}

\title{A steady solution to the hydrodynamic equation and incommensurate magnetization in  a U(2) invariant superfluid}

%\author{\author{Yi-Cai Zhang}{zhangyicai123456@163.com}}
%\address{School of Physics and Materials Science, Guangzhou University, Guangzhou 510006, People's Republic of China }
%
%\author{\author{Shizhong Zhang}{shizhong@hku.com}}\address{Department of Physics and Hong Kong Institute of Quantum Science and Technology, The University of Hong Kong, Hong Kong, China}

\author{Guang-Xin Pang$^1$, Yi-Cai Zhang$^1$}
\address{$^1$ School of Physics and Materials Science, Guangzhou University, Guangzhou 510006, People's Republic of China}

%\address{$^2$ Department of Physics and Hong Kong Institute of Quantum Science and Technology, The University of Hong Kong, Hong Kong, China}
%\ead{shizhong@hku.hk}
\ead{zhangyicai123456@163.com}

\vspace{10pt}
%\begin{indented}
%\item[]August 2017 (minor update March 2024)
%\end{indented}

\begin{abstract}
At the zero temperature limit, a one-dimensional steady solution to the hydrodynamic equation of a U(2) invariant superfluid is obtained. This solution reveals that the magnitude of magnetization is always directly proportional to the particle number density. Furthermore, the problem can be interpreted as a particle's motion in a central force field. It is demonstrated that the particle's orbits are elliptical in shape, with a precession angle determined by a non-zero mass current. This suggests that the spatial periods of the three component magnetizations are not commensurate. These findings indicate that the coupling of mass superflow and magnetization distortions usually results in an incommensurate magnetization.

\end{abstract}

%\maketitle

%\maketitle
\section{Introduction}
 The low-energy dynamics of a fluid can be described by hydrodynamic equation. In a classical ideal  fluid, the hydrodynamic variables usually refer to the densities of conserved quantities, for example, the density of particle number, the momentum density, the entropy density\cite{Forster}.  When symmetry breaking occurs, additional hydrodynamic variables are introduced, such as  the superfluid velocity in usual simple superfluid system (e.g., superfluid $^4$He or single component Bose-Einstein condensate of atomic gas) \cite{Chaikin}.
Recently, the physics of SU(N) atomic gas has attracted a lot of interest \cite{Sonderhouse,Taie, Cazalilla, Ren, Cai}. The SU(N) symmetry may lead to unconventional magnetism \cite{Honerkamp},
spin-liquid \cite{Yu}, superfluid \cite{Cherng}, itinerant ferromagnetism \cite{PeraJordi}, valence-bond solid phase \cite{XuYu}, trion states \cite{Han} , etc.
The collective excitation \cite{Pagano,He} and the equation of state  \cite{Pasqualetti} of SU(N) gases have been measured experimentally.
  Mott crossover \cite{Hofrichter} and bosonization of fermion \cite{Song} are observed experimentally.
The hydrodynamics are expected to provide necessary tools for the investigations of the collective dynamics of SU (N) atomic gases.

In a previous work by one of the authors\cite{yicai2024}, with the Hamilton method, the hydrodynamic equation has been generalized to a superfluid with U(N) invariant interaction.
In this paper, based on the hydrodynamic equation,
we will investigate a one-dimensional steady solution in a U(2) invariant superfluid at zero temperature.

Our findings show that the solution to this problem can be achieved by mapping it onto a particle's motion in an effective central force field. When the mass current is zero, the particle's orbit is closed, indicating that the spatial periods of the three-component magnetization are commensurate. However, when the mass current is non-zero, the particle's orbit is no longer closed and instead forms an ellipse with a non-zero precession angle. In this scenario, the spatial periods of the three magnetization components are not commensurate in general. This demonstrates the influence of the coupling between superflow and magnetization, as a mass superflow can induce interesting spatial structures in the magnetization.

\section{Hydrodynamic equation for a U(N) invariant superfluid}
In a superfluid, the fluid can be divided into two distinct parts: the normal part with a particle number density of $n_n$, and the superfluid part with a particle number density of $n_s$. The total particle number density is the sum of these two parts, denoted as $n = n_n + n_s$. Now the liquid can take part in two kinds of motions \cite{Tisza,Landau,Lifshitz,zhangyicai2019}.
The superfluid part would have a velocity $\mathbf{v}_s$ and the normal part would have a velocity $\mathbf{v}_n$.

The  thermodynamic relations of a superfluid with U(N) invariant interaction  takes the form of \cite{yicai2024}
\begin{eqnarray}
&d\epsilon=Tds+\mu dn+\mathbf{v}_{n}\cdot d\mathbf{g}+\mathbf{h}\cdot d\mathbf{v}_{s}+\mu^adn^a,\nonumber\\
&f=\epsilon-Ts-\mathbf{v}_n\cdot\mathbf{g},\nonumber\\
&df=-sdT+\mu dn-\mathbf{g}\cdot d\mathbf{v}_n+\mathbf{h}\cdot d
\mathbf{v}_{s}+\mu^adn^a,\nonumber\\
& p=-\epsilon +Ts+\mathbf{v}_{n}\cdot \mathbf{g}+\mu n+\mu^{a}n^a,
 \nonumber\\
&d p= sdT+\mathbf{g}\cdot d\mathbf{v}_n+nd\mu -\mathbf{h}\cdot d
\mathbf{v}_{s}+n^ad\mu^a,\label{29}
\end{eqnarray}
where $\epsilon$ is energy density (energy per unit volume), $f$ is free energy density (free energy per unit volume), $p$ is pressure. $T$ is temperature, $s$ is entropy density (entropy per unit volume), $\mu$ is chemical potential and $n$ is the particle number density (particle number per unit volume).
$\mathbf{v}_n$ is the velocity of normal part, $\mathbf{g}$ is momentum density (momentum per unit volume),  $\mathbf{v}_s$ is the velocity of superfluid part, and
\begin{eqnarray}
\mathbf{h}=n_{s}m(\mathbf{
v}_{s}-\mathbf{v}_{n})
\end{eqnarray}
is the conjugate variable of superfluid velocity $\mathbf{v}_s$ and $m$ is the mass of a particle  in fluid.
$n^a$ is density for  conserved SU(N) charge (generalized spin \cite{zhangyicai2018}) and
$\mu^a$ are generalized chemical potential for conserved charge $n^a$.
 Here the hydrodynamic variables are taken as entropy density $s$, the particle number density $n$ [U(1) charge], the momentum density $\mathbf{g}$, the superfluid velocity $\mathbf{v}_s$, and the conserved SU(N)
  charge (generalized spin) density $n^a$.

With the Hamilton method and a set of commutation  relations (Poisson's brackets) among these hydrodynamic variables, we derive the  hydrodynamic  equation for a U(N) invariant superfluid \cite{yicai2024}, i.e.,
\begin{eqnarray}
& \frac{\partial n}{\partial t}+\nabla \cdot \mathbf{j}=0,\label{continuity11}\\
& \frac{\partial g_{i}}{\partial t}+\sum_{j}\partial _{j}\pi_{ij}=0, \label{continuity12}\\
& \frac{\partial s}{\partial t}+\nabla\cdot(s\mathbf{v}_n)=0,\label{continuity13}\\
& \frac{\partial n^a}{\partial t}+\nabla \cdot \mathbf{j}^{a}=0,\label{47}\\
&\frac{\partial \mathbf{v}_{s}}{\partial t}+\nabla (\frac{\mu}{m}+\mathbf{v}_n \cdot \mathbf{v}_s)=-(\nabla\times \mathbf{v}_s)\times \frac{n_n\mathbf{v}_n+n_s \mathbf{v}_s}{n}\nonumber\\
&-\frac{n^a(\nabla \mu^a)}{mn}, \label{continuity112}
\end{eqnarray}
with constitutive relations
\begin{eqnarray}
& j_{i}=n_{n}\textrm{v}_{ni}+n_{s}\textrm{v}_{si}, \\
& g_{i}=mj_{i}, \\
& \pi _{ij}=p\delta _{ij}+mn_{n}\textrm{v}_{ni}\textrm{v}_{nj}+mn_{s}\textrm{v}_{si}\textrm{v}_{sj}\\
&\mathbf{j}^{a}=\alpha f^{c}_{ab}n^c\nabla n^{b}+\frac{n^a(n_n\mathbf{v}_n+n_s\mathbf{v}_s)}{n},\label{111}
%&=\alpha f^{c}_{ab}n^c\nabla n^{b}+n^a\mathbf{\bar{v}}.
\end{eqnarray}
where indices $i,j=x,y,z$,   $\alpha$ is a constant, $\mathbf{j}$ is particle current density, $m$ is particle's mass in the liquid, and $f^{c}_{ab}$ is the structure constant of the SU(N) Lie algebra of the U(N) group. This structure constant is completely antisymmetric with respect to indices (abc). In the case of U(2), $f^{c}_{ab}$ is the Levi-Civita tensor, which can be represented as $\epsilon_{abc}$ and has a value of $\epsilon_{123}=1$. In the Eq. (\ref{111}) above, we have used Einstein's summation convention for the indices $a, b, c$. Eqs. (\ref{continuity11}), (\ref{continuity12}), (\ref{continuity13}), and (\ref{47}) respectively describe the conservation of particle number, momentum, entropy, and SU(N) charge (generalized spin). Eq. (\ref{continuity112}) represents the equation of motion for the superfluid velocity $\mathbf{v}_s$.

Due to the coupling between generalized spin (magnetization) and superflow, the motion of magnetization $n^a$ can be affected by a mass superflow. In the next section, we will present a steady solution to the above equations. It will be demonstrated that the magnitude of magnetization, denoted as $M=\sqrt{n^an^a}$, is always directly proportional to the particle number density $n$. Additionally, our findings indicate that a nonzero particle current $\mathbf{j}\neq0$ (or mass superflow), typically induces an incommensurate magnetization structure.

\section{An one-dimensional steady solution at absolute zero }\label{sec:4}
In a one-dimensional steady superflow scenario, when the temperature is  at $T=0$, the entire liquid becomes a full superfluid. This means that $n_s=n$,  and $n_n=v_n=s=0$. For the purposes of this discussion, we will assume that the fluid is flowing in the x-direction. As a result, the hydrodynamic Eqs. (\ref{continuity11})-(\ref{continuity112}) can be simplified to
\begin{eqnarray}
\label{continuity1}
& \frac{\partial n}{\partial t}+\partial_x(nv_s)=0,
 \\
 \label{continuity2}
& \frac{\partial(nmv_s)}{\partial t}+\partial_x(p+mnv_{s}^2)=0, \\
 \label{continuity4}
&\frac{\partial v_{s}}{\partial t}+\frac{\partial_x \mu}{m}+\frac{n^a(\partial_x \mu^a)}{mn}=0,\\
 \label{continuity3}
& \frac{\partial n^a}{\partial t}+\partial_x(\alpha f^{a}_{bc}n^c\partial_x n^{b}+n^av_s)=0,
\end{eqnarray}%
where $v_s\equiv v_{sx}$, $j\equiv j_x=nv_s$, $g=g_x=mj=nmv_s$, and $\pi_{xx}=p+mnv_{s}^2$.

Using the thermodynamic relation for $T=0$ [the generalized Gibbs-Duhem relation, Eq.(\ref{29})]  i.e., $dp=nd\mu +n^ad\mu^a-mnv_s d
v_{s}$ and continuity Eq.(\ref{continuity1}), one can show that Eq.(\ref{continuity2}) and (\ref{continuity4}) are equivalent, and then Eq.(\ref{continuity4}) can be omitted. So the above equation is  further reduced into
\begin{eqnarray}
& \frac{\partial n}{\partial t}+\partial_x(nv_s)=0,
\label{continuity} \\
& \frac{\partial(nmv_s)}{\partial t}+\partial_x(p+mnv_{s}^2)=0, \\
& \frac{\partial n^a}{\partial t}+\partial_x(\alpha f^{a}_{bc}n^c\partial_x n^{b}+n^av_s)=0.
%&\frac{\partial v_{s}}{\partial t}+\partial_x \mu+\frac{n_a(\partial_x \mu_a)}{n}=0
\end{eqnarray}%

In the following, we seek a steady solution to  the equation mentioned above and
 the time's dependence are neglected. This will allow us to obtain a set of integral constants, namely:
 \begin{eqnarray}
&J_0\equiv j=nv_s , \label{630}\\
&p_0\equiv p+mnv_{s}^2, \label{63}\\
&C^\alpha \equiv \alpha f^{a}_{bc}n^c\partial_x n^{b}+n^av_s.\label{55}
%&\frac{\partial v_{s}}{\partial t}+\partial_x \mu+\frac{n_a(\partial_x \mu_a)}{n}=0
\end{eqnarray}
%where $J_0, p_0, C^{a}$ are integral constants.
By utilizing the complete antisymmetry of the structure constant $f_{bc}^{a}$, we can derive the following equations from Eq.(\ref{630}) and (\ref{55}):
\begin{eqnarray}
&\frac{J_0 n^a n^a}{n}=C^a n^a,\nonumber\\
&\frac{J_0 n^a \partial_x n^a}{n}=C^a \partial_x n^a,
\label{56}
\end{eqnarray}
where the repeated indices $a$ imply summation.
 Solving  Eq.(\ref{56}) , we get
\begin{eqnarray}\label{Dn1}
& \sqrt{n^a(x) n^a(x)}=D n(x),\nonumber\\
&J_0 D^2n(x)=C^a n^a(x),
\end{eqnarray}
where $D\geq0$ is also an integral constant.  Eq.(\ref{Dn1}) shows that the magnitude of generalized spin density $\sqrt{n^a n^a}$ is always proportional to particle number density $n$. We should emphasize that this result holds in a general U(N) invariant superfluid, not only limited to the U(2) case.

For a U(2) invariant superfluid, the conserved charge density $n^a$ is the magnetization $M^{x,y,z}$.
Then the equation for magnetization is
\begin{eqnarray}\label{Dn}
& M(x)=\sqrt{M^2(x)}=D n(x),\nonumber\\
&J_0 D^2n(x)=C^aM^a(x),
\end{eqnarray}
where $M^2(x)=(M^x)^2+(M^y)^2+(M^z)^2$.

Because $C^{a}=(C^x,C^y,C^z)$ can be viewed as a vector in the three dimension (internal spin) space, then without loss of generality,
in the following we would  assume
\begin{eqnarray}\label{CC}
C^x=C^y=0, \ \ \ C\equiv C^z\neq0.
\end{eqnarray} Based on Eq.(\ref{Dn}), then we get
\begin{eqnarray}\label{61}
&M^z(x)=\frac{J_0 D^2n(x)}{C}=D cos(\theta) n(x),
\end{eqnarray}%
where we introduce a constant polar angle $\theta$ by
\begin{eqnarray}\label{2311}
cos(\theta)\equiv \frac{J_0 D}{C}.
\end{eqnarray}%
 Due to constraint of $0\leq| cos(\theta)|\leq 1$, current $J_0$ satisfies relation
 \begin{eqnarray}\label{J0}
0\leq |J_0|\leq |C/D|.
\end{eqnarray}%
By Eq.(\ref{Dn}), the other two magnetization components can be represented by
\begin{eqnarray}\label{621}
&M^x=Dsin(\theta)n(x) cos(\phi(x)),\nonumber\\
&M^y=Dsin(\theta)n(x) sin(\phi(x)),
\end{eqnarray}
where the polar angle $0\leq\theta\leq\pi$ and azimuthal angle $0\leq\phi\leq2\pi$ in spherical coordinate system of three dimension (spin) space.

By Eq.(\ref{630}), Eq.(\ref{55}), Eq.(\ref{61}) , Eq.(\ref{2311}) and Eq.(\ref{621}), we get
\begin{eqnarray}\label{800}
&\alpha[ M^y\partial_x M^{x}-M^x\partial_x M^{y}]=C-\frac{J_0 M^z}{n}=C-J_0D cos(\theta).\nonumber\\
&\Rightarrow\frac{d\phi}{dx }=\frac{J_0D cos(\theta)-C}{\alpha D^2 sin^2(\theta) n^2(x)}=\frac{-\kappa}{ n^2(x)},
\end{eqnarray}
where we introduce an constant
\begin{eqnarray}
\kappa\equiv \frac{-J_0D cos(\theta)+C}{\alpha D^2 sin^2(\theta)}=\frac{C}{\alpha D^2}.
\end{eqnarray}
With Eq.(\ref{CC}),  Eq.(\ref{61}), Eq.(\ref{621}) and Eq.(\ref{800}), one can verify that the $x-$ and $y-$ component magnetization $M^x$, $M^y$ satisfy  Eq.(\ref{55}) automatically.

To further determine the density function $n(x)$, it is necessary to have knowledge of either the energy function $\epsilon(n,M^a,v_s)$ or the pressure function $p(n,M^a,v_s)$ (also known as the equation of state) \cite{fluid}. In the following, we will assume that the pressure function is represented by the following equation:
\begin{eqnarray}\label{79}
%&E=\int dx \epsilon \\
&p(n,M^a,v_s)=\frac{g_0n^2}{2}+\frac{g_2 M^2}{2}+\frac{\alpha (\partial _x M^a)^2}{2}
\end{eqnarray}
where $g_{0}, g_{2}$ are density-density and spin-spin interaction constants between atoms, respectively. The above form of pressure function $p$ is relevant to the dilute cold atomic gas \cite{yicai2024,Ueda}.
Then  the constant $p_0$ in Eq.(\ref{63}) is
\begin{eqnarray}\label{581}
&p_0=p+mnv^{2}_{s}=\frac{g_0n^2}{2}+\frac{g_2 M^2}{2}+\frac{\alpha (\partial _x M^a)^2}{2}+mn v_{s}^2.
\end{eqnarray}
Using Eqs.(\ref{630}), (\ref{61}), (\ref{621}),  and (\ref{800}), we get
\begin{eqnarray}
&p_0=\frac{1}{2}[\alpha D^2(\partial_x n)^2+\frac{\alpha \kappa^2 D^2sin^2(\theta)}{n^2}]+\frac{mJ_{0}^2}{n}+\frac{(g_0+g_2D^2)n^2}{2}.
\end{eqnarray}
Furthermore we introduce a reduced constant $\tilde{p}_0$ by
\begin{eqnarray}\label{62}
&\tilde{p}_0\equiv p_0/(\alpha D^2)=\frac{1}{2}(\partial_x n)^2+\frac{\beta}{2n^2}+\frac{\gamma}{n}+\frac{\delta}{2} n^2,
\end{eqnarray}
where \begin{eqnarray}\label{281}
&\beta\equiv\kappa^2sin^2(\theta)=\frac{C^2-J_{0}^{2}D^2}{\alpha^2D^4}\geq0,\nonumber\\
&\gamma\equiv\frac{mJ^{2}_{0}}{\alpha D^2}\geq0,\nonumber\\
&\delta\equiv\frac{g_0+g_2D^2}{\alpha D^2}.
\end{eqnarray}
We find that if the interaction parameter $\delta\leq0$, the particle density $n(x)$ will increase without bound as $x$ approaches infinity, indicating instability in the system. In order to obtain a finite solution for $n(x)$, we will assume that the interaction parameter $\delta>0$.
By Eq.(\ref{62}), then we get
\begin{eqnarray}
\label{87}
&\frac{dn}{dx}=\sqrt{2[\tilde{p}_0-(\frac{\beta}{2n^2}+\frac{\gamma}{n}+\frac{\delta }{2}n^2)]}\nonumber\\
&\Rightarrow x-x_0=\int\frac{dn}{\sqrt{2[\tilde{p}_0-V(n))]}},
\end{eqnarray}
where
\begin{eqnarray}
V(n)=\frac{\beta}{2n^2}+\frac{\gamma}{n}+\frac{\delta }{2}n^2.
\end{eqnarray}
%where we take initial condition  that when $x=x_0$, $n(x=x_0)=n(x_0)$.
In general, this integral expression Eq.(\ref{87}) can be represented by the elliptic integrals (and elliptic functions) \cite{Abramowitz,Wang}. Due to the
complicated  expressions of elliptic integrals, we will integrate Eq.(\ref{87}) directly by numerical method, rather write them out explicitly. Further using Eqs.(\ref{800}) and (\ref{87}), we get an equation for $\phi$
\begin{eqnarray}
\label{94}
&\frac{d\phi}{dn}=\frac{-\kappa}{n^2\sqrt{2[\tilde{p}_0-V(n)]}}\nonumber\\
&\Rightarrow \phi-\phi_0=\int\frac{-\kappa dn}{n^2\sqrt{2[\tilde{p}_0-V(n)]}}.
\end{eqnarray}

%\begin{verbatim}
%\includegraphics{fig0.eps}
%\end{verbatim}\normalsize

\begin{figure}[th]
\centering
\includegraphics[scale=0.45]{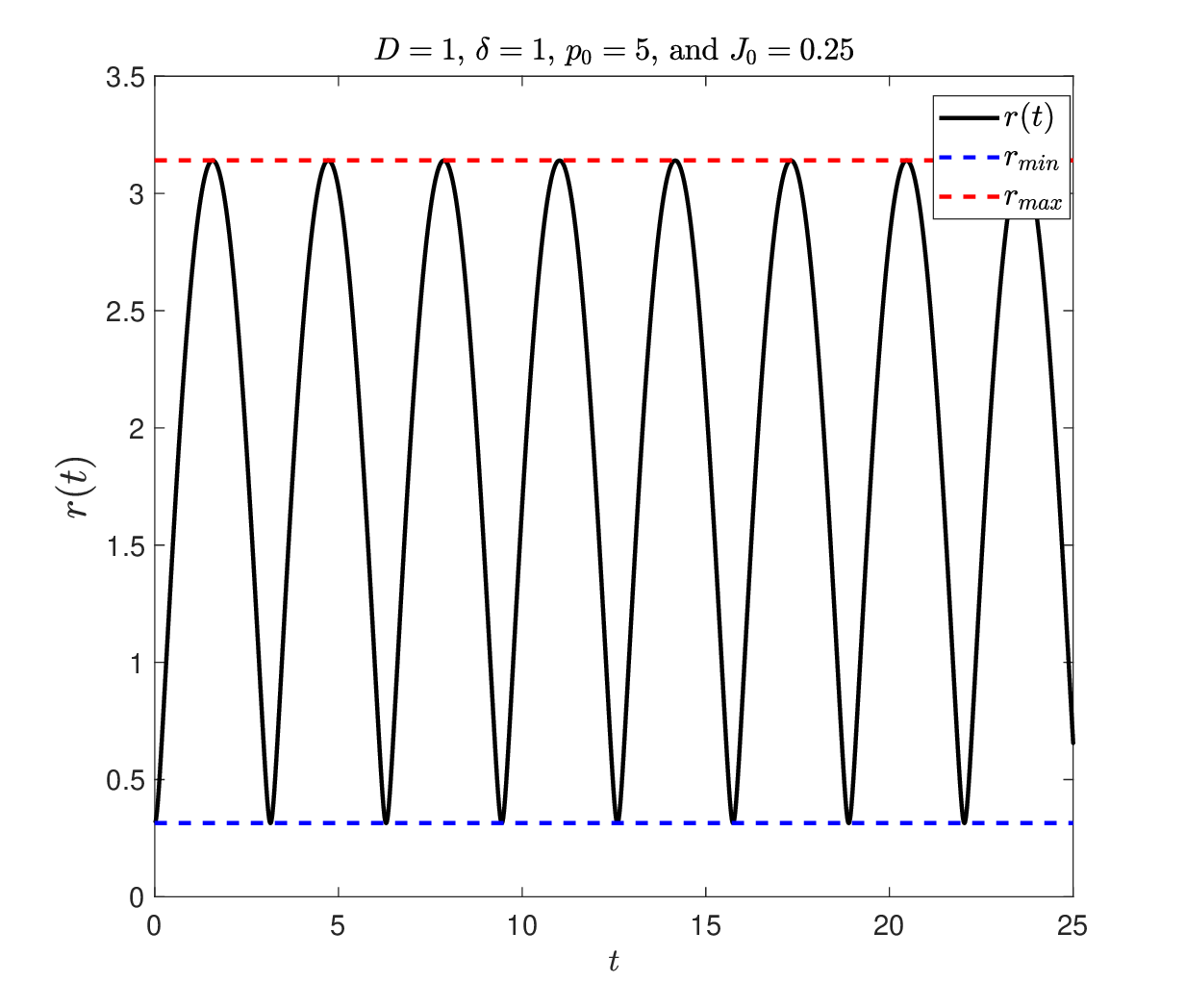}
\caption{The distance $r(t)$ of a particle from origin (force center) as a periodic function of time $t$ with parameters $D=1$, $p_0=5$ and $J_0=0.25$.}
\label{fig:1}
\end{figure}

\begin{figure}[th]
\centering
\includegraphics[scale=0.45]{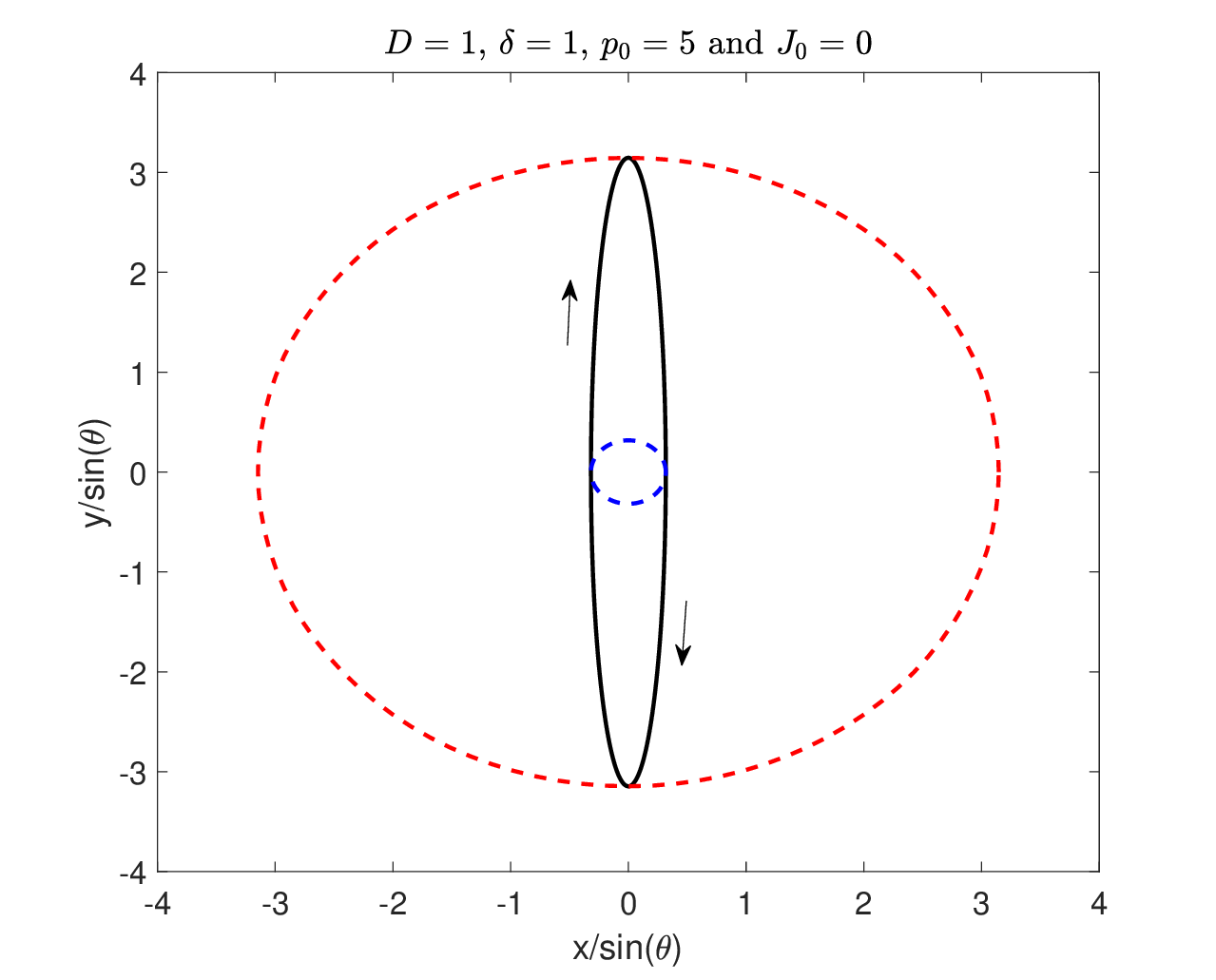}
\caption{The orbit of the $x-y$ plane with $p_0=5$,  $J_0=0$ and $sin(\theta)=1$. is a closed Hooke's ellipse with a center at the origin (force center). The arrows indicate the direction of the particle's motion. The radius of the large (small) circle is $r_{max}$ ($r_{min}$).  }
\label{fig:1}
\end{figure}

\begin{figure}[th]
\centering
\includegraphics[scale=0.45]{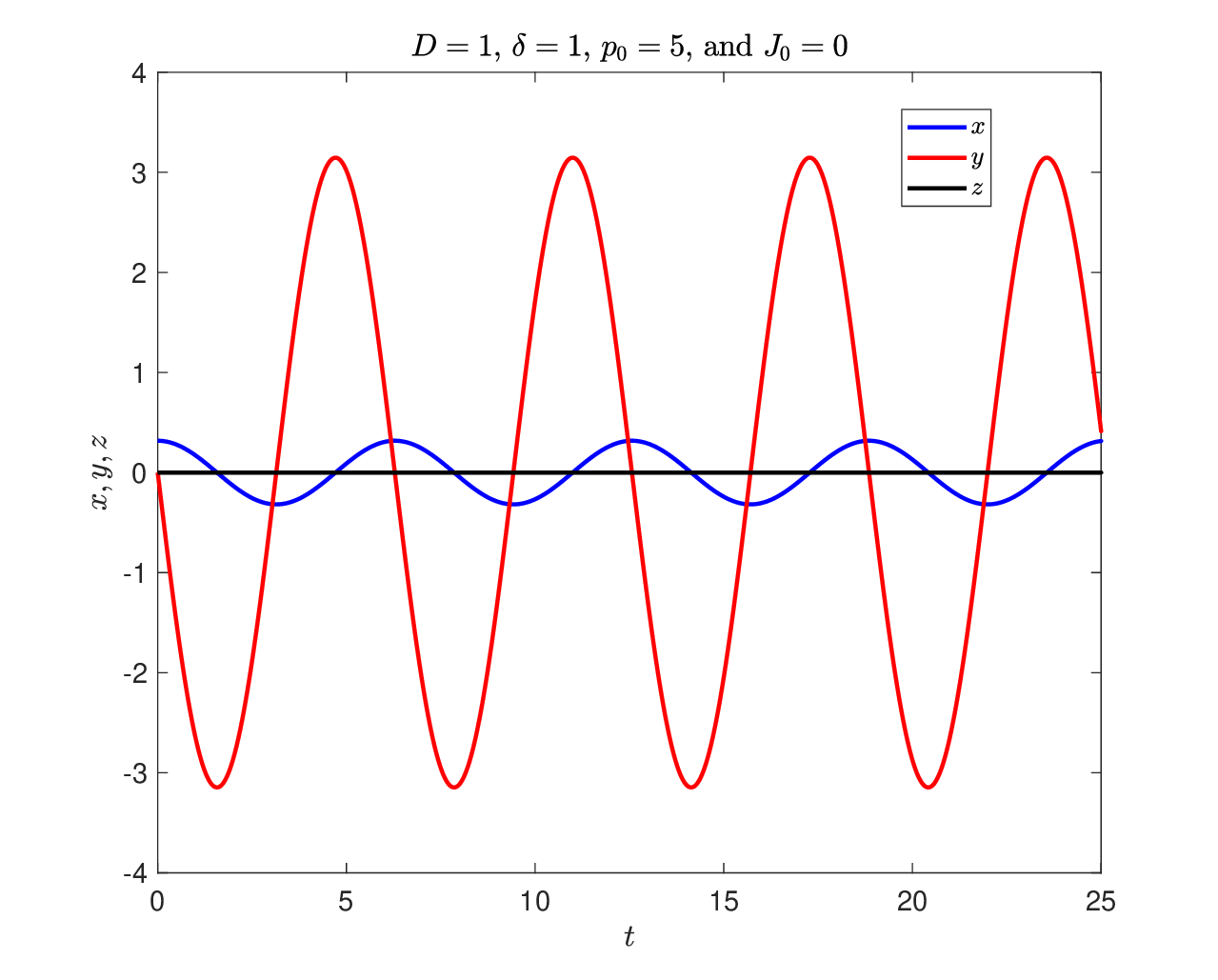}
\caption{ $x,y,z$ coordinates as functions of time $t$. The variations of $x,y$ with time corresponds to the orbit of Fig.2. When $J_0=0$, the $z-$ component magnetization is always zero (see the black line). The periods of $x$ and $y$ are the same [see Eq.(\ref{89})], and the corresponding orbit is closed (see Fig.2).}
\label{fig:1}
\end{figure}

In the following, we would use an analogy of classical mechanics to reformulate the above problem. The original one-dimension spatial coordinate $x$ is identified as time $t$, and the three magnetization component $M^{x,y,z}/D$ can be viewed as the three coordinate component $x,y,z$ in three dimension space and $n$ is identified with the distance of particle  from the origin $r$, i.e,
 \begin{eqnarray}\label{791}
&x\rightarrow t, \nonumber\\
&n\rightarrow r=\sqrt{x^2+y^2+z^2},\nonumber\\
&M^x/D\rightarrow  x=rsin(\theta)cos(\phi),\nonumber\\
&M^y/D\rightarrow  y=rsin(\theta)sin(\phi),\nonumber\\
&M^z/D\rightarrow  z=rcos(\theta),\nonumber\\
&M^2/D^2=n^2\rightarrow r^2=x^2+y^2+z^2,
\end{eqnarray}
where the polar angle $0\leq\theta\leq\pi$ and azimuthal angle $0\leq\phi\leq2\pi$.
Then, in the picture of particle's motion, Eq.(\ref{800}) is conservation of z-component angular momentum, Eq.(\ref{87}) is radial equation of  motion for particle in a central force field with an effective potential
\begin{eqnarray}\label{80}
V(n)\rightarrow V(r)=\frac{\beta}{2r^2}+\frac{\gamma}{r}+\frac{\delta }{2}r^2,
\end{eqnarray}
and Eq. (\ref{94}) represents the orbit equation, while $\tilde{p}_0$ can be interpreted as the total mechanical energy in Eq.(\ref{62}). When considering specific values for the parameters $J_0$, $D$, $C$, and  then by Eq.(\ref{2311}), $\theta$ takes  a fixed value. This can be seen as a problem of a particle moving on a right circular   cone surface (polar angle $\theta= Const$) with its vertex at the origin (force center).

 In the effective potential Eq. (\ref{80}), the first term $\frac{\beta}{2r^2}$ represents the centrifugal potential arising from angular momentum conservation, the second term $\frac{\gamma}{r}$ represents the repulsive Coulomb potential ($\gamma\geq0$), and the last term $\frac{\delta }{2}r^2$ represents the harmonic oscillator potential. According to Eq. (\ref{62}), for a specific total energy $\tilde{p}_0$, when the kinetic energy in the radial direction becomes zero (i.e. $\frac{1}{2}(\partial_t r)^2=0$), the energy equation
 \begin{eqnarray}\label{En} &\tilde{p}_0=p_0/(\alpha D^2)=\frac{\beta}{2r^2}+\frac{\gamma}{r}+\frac{\delta}{2} r^2, \end{eqnarray}
 typically has two real roots: $r=r_{min}$ and $r=r_{max}$. The distance of the particle from the origin, $r(t)$, will vary between $r_{min}$ and $r_{max}$, meaning that $r_{min}\leq r\leq r_{max}$. This distance will be a periodic function of time $t$ (see Fig.1).

In the following, we will use the system of units where $m=\alpha=C=1$. Additionally, we will set some fixed parameters as $D=1$, $\delta=1$, and $p_0=5$. In Fig.1, we have chosen $J_0=0.25$ to plot the function $r(t)$ as a function of time $t$. It is evident that $r(t)$ is a periodic function of time $t$, which implies that the density $n(x)$ is also a periodic function of the coordinate $x$. However, this does not necessarily mean that the orbit of the particle in three-dimensional space is closed. According to Bertrand's theorem in classical mechanics \cite{Arnold}, the orbit of a particle is closed only when the potential is either the attractive Coulomb potential or the harmonic oscillator potential. In the case where the repulsive Coulomb potential exists in Eq.(\ref{80}), i.e., when $\gamma\neq0$ ( i.e., $J_0\neq0$), the orbit of the particle would generally not be closed. In the following, we will discuss two cases: $J_0=0$ and $J_0 \neq 0$.

\begin{figure}[t]
%æ¯å¯éé¡¹ hè¡¨ç¤ºçæ¯hereå¨è¿éæå¥ï¼tè¡¨ç¤ºçæ¯å¨é¡µé¢çé¡¶é¨æå¥
\centering
\includegraphics[scale=0.43]{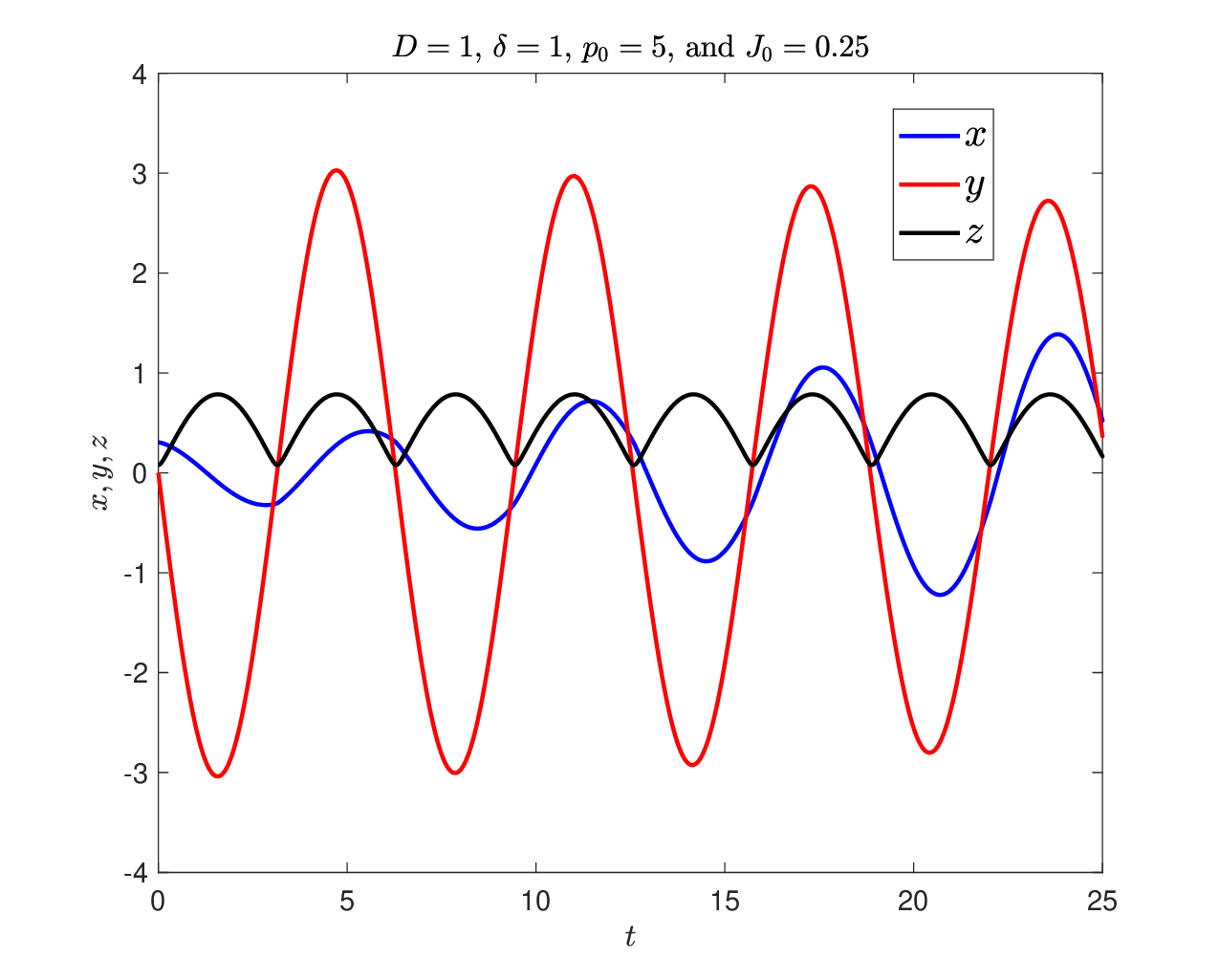}
\caption{The coordinates of the particle, $x$, $y$, and $z$, are functions of time $t$. The variations of $x$ and $y$ over time correspond to the orbit shown in Fig.5. In this figure, $J_0$ is equal to 0.25 and the $z$-component of magnetization is not equal to zero, as indicated by the black line. As time increases, the value of $x$  increases while the value of $y$ decreases. This suggests that the orbit shown in Fig.5 is not closed and the periods of $x$ and $y$ are not commensurate.}
\label{fig:1}
\end{figure}

\begin{figure}[th]
%æ¯å¯éé¡¹ hè¡¨ç¤ºçæ¯hereå¨è¿éæå¥ï¼tè¡¨ç¤ºçæ¯å¨é¡µé¢çé¡¶é¨æå¥
\centering
\includegraphics[scale=0.45]{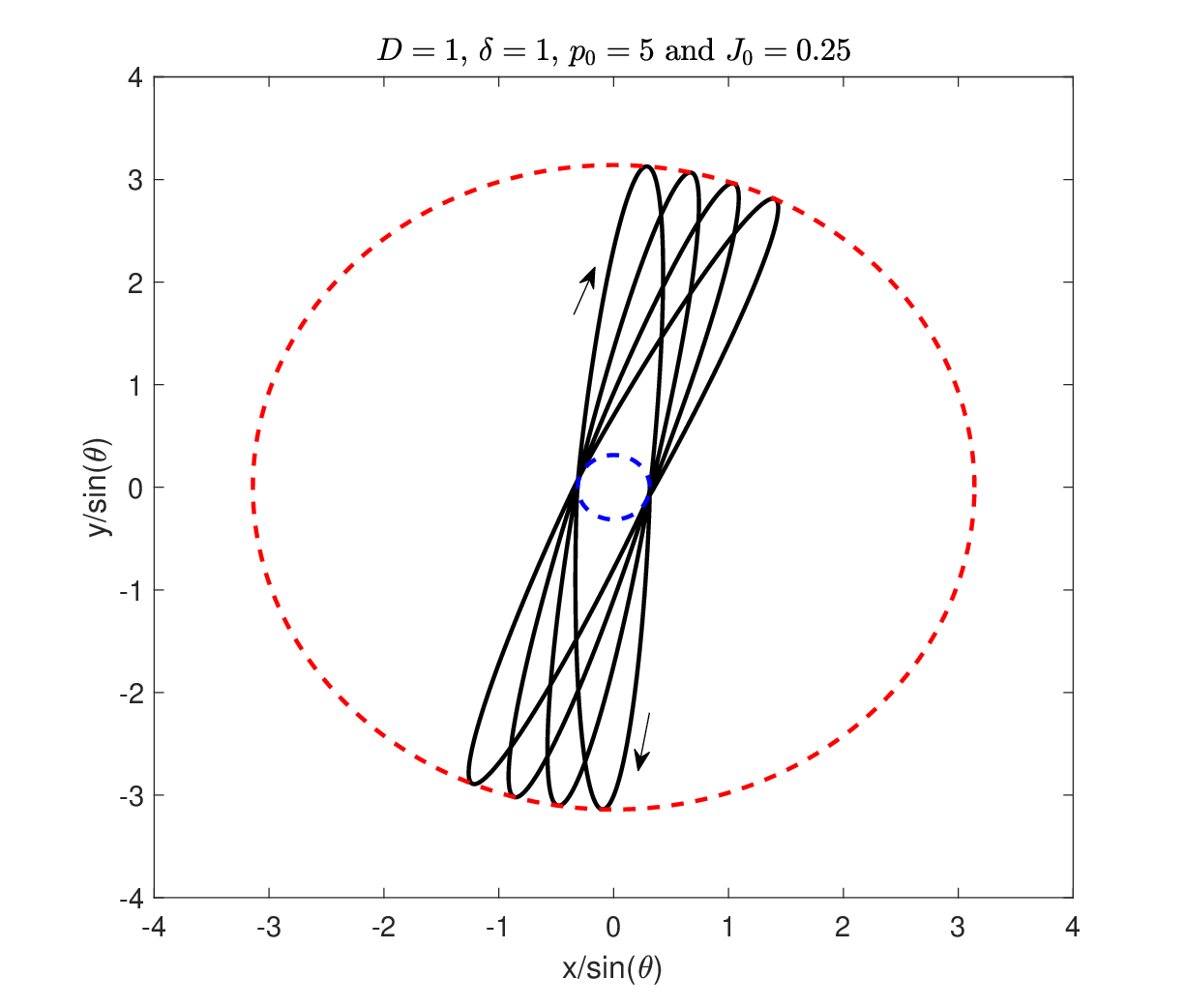}
\caption{The orbit of the $x-y$ plane with $p_0=5$,  $J_0=0.25$, and $sin(\theta)=0.968$. This can be described as a Hooke's ellipse with a slow precession. The arrows indicate the direction of the particle's motion.}
\label{fig:1}
\end{figure}

\subsection{$J_0=0$ (closed Hooke's ellipses) }
When $J_0=0$, by Eq.(\ref{61}) and (\ref{281}) , then $M^z=0$, $\theta=\pi/2$, $\beta=\kappa^2=\frac{C^2}{\alpha^2 D^4}$, $\gamma=0$. In such a case, the second term $\frac{\gamma}{r}$ in Eq. (\ref{80}) is absent, then the particle moves in  a three dimension harmonic oscillator potential. The configuration space of particle's motion  is a two dimensional $x-y$ plane.  According to Bertrand's theorem in classical mechanics, the orbit of particle would be closed Hooke's ellipse.
 In specific, by Eq.(\ref{87}), we have
\begin{eqnarray}
&t-t_0=\int\frac{dr}{\sqrt{2\{\tilde{p}_0-[\frac{\kappa^2}{2r^2}+\frac{\delta}{2} r^2]\}}}=\frac{1}{2\sqrt{\delta}}Arcsin[\frac{r^2-\frac{\tilde{p}_0}{\delta}}{\sqrt{\frac{\tilde{p}_{0}^{2}}{\delta^2}-\frac{\kappa^2}{\delta}}}].
\end{eqnarray}
Furthermore, considering the initial condition that when $t=0$, the particle is at the nearest point apart from force center (the origin), i.e., ``perihelion", then, $r^2(t=0)=r^{2}_{min}=\frac{\tilde{p}_0}{\delta}-\sqrt{\frac{\tilde{p}_{0}^{2}}{\delta^2}-\frac{\kappa^2}{\delta}}$ and $t_0=\pi/(4\sqrt{\delta})$, we get an equation of $r(t)$, i.e.,
\begin{eqnarray}\label{351}
r^2(t)=\frac{\tilde{p}_0}{\delta}-\sqrt{\frac{\tilde{p}_{0}^2}{\delta^2}-\frac{\kappa^2}{\delta}}cos(2\sqrt{\delta}t).
\end{eqnarray}

By Eqs.(\ref{800}) and (\ref{351}), we get angle $\phi$
\begin{eqnarray}
&\phi-\phi_0=-\int\frac{\kappa dt}{\frac{\tilde{p}_0}{\delta}-\sqrt{\frac{\tilde{p}_{0}^2}{\delta^2}-\frac{\kappa^2}{\delta}}cos(2\sqrt{\delta}t)}\nonumber\\
&=-\frac{\kappa Arctan[\frac{\tilde{p}_0+\sqrt{\tilde{p}_{0}^2-\delta\kappa^2}}{\sqrt{\delta\kappa^2}}tan(\sqrt{\delta}t)]}{|\kappa|}.
\end{eqnarray}
Taking an initial condition $t=0$, $\phi(t=0)=0$, we get
\begin{eqnarray}\label{371}
\phi=\frac{-\kappa Arctan[\frac{\tilde{p}_0+\sqrt{\tilde{p}_{0}^2-\delta\kappa^2}}{\sqrt{\delta\kappa^2}}tan(\sqrt{\delta}t)]}{|\kappa|}.
\end{eqnarray}

Using Eq.(\ref{94}), the orbit equation is reduced into
\begin{eqnarray}
&\phi-\phi_0=-\int\frac{\kappa dr}{r^2\sqrt{2\{\tilde{p}_0-[\frac{\kappa^2}{2r^2}+\frac{\delta}{2} r^2]\}}}\nonumber\\
&=\frac{\kappa}{2|\kappa|}Arcsin[\frac{\frac{1}{r^2}-\frac{\tilde{p}_0}{\kappa^2}}{\sqrt{\frac{\tilde{p}_{0}^{2}}{\kappa^4}-\frac{\delta}{\kappa^2}}}].
\end{eqnarray}
Further when we take initial condition that when $t=0$, $\phi(t=0)=0$, and $1/r^2(t=0)=1/r_{min}^2=\frac{\tilde{p}_0}{\kappa^2}+\sqrt{\frac{\tilde{p}_{0}^{2}}{\kappa^4}-\frac{\delta}{\kappa^2}}$, we get the obit equation
\begin{eqnarray}
1/r^2=\frac{\tilde{p}_0}{\kappa^2}+\sqrt{\frac{\tilde{p}_{0}^2}{\kappa^4}-\frac{\delta}{\kappa^2}}cos[\frac{2|\kappa|\phi}{\kappa})].
\end{eqnarray}
Due to $cos(2\phi)=cos(-2\phi)=cos^2(\phi)-sin^2(\phi)$, $z=0$ and $r^2=x^2+y^2$, then, we get the orbit equation \begin{eqnarray}
&1=r^2\frac{\tilde{p}_0}{\kappa^2}+\sqrt{\frac{\tilde{p}_{0}^2}{\kappa^4}-\frac{\delta}{\kappa^2}}r^2cos[2\phi]\nonumber\\
&\Rightarrow\frac{\tilde{p}_0}{\kappa^2}(x^2+y^2)+\sqrt{\frac{\tilde{p}_{0}^2}{\kappa^4}-\frac{\delta}{\kappa^2}}(x^2-y^2)=1.
\end{eqnarray}
We see that the orbit   is a  Hooke's ellipse with center at the origin (the force center).
Further using Eqs.(\ref{351}) and (\ref{371}), the coordinate component $x,y,z$ are given by
 \begin{eqnarray}
&x=rsin(\theta)cos(\phi)=\sqrt{\frac{\tilde{p}_0}{\delta}-\sqrt{\frac{\tilde{p}_{0}^2}{\delta^2}-\frac{\kappa^2}{\delta}}}cos(\sqrt{\delta}t),\nonumber\\
& y=rsin(\theta)sin(\phi)=-\frac{\kappa}{|\kappa|}\sqrt{\frac{\tilde{p}_0}{\delta}+\sqrt{\frac{\tilde{p}_{0}^2}{\delta^2}-\frac{\kappa^2}{\delta}}}sin(\sqrt{\delta}t),\nonumber\\
& z=rcos(\theta)=0,
\end{eqnarray}
where $\theta=\pi/2$.
Transforming them  back the original physical quantities, then the particle number density and magnetization are
\begin{eqnarray}\label{89}
&n(x)=r=\sqrt{\frac{\tilde{p}_0}{\delta}-\sqrt{\frac{\tilde{p}_{0}^2}{\delta^2}-\frac{\kappa^2}{\delta}}cos(2\sqrt{\delta}x)},\nonumber\\
&M^x(x)=Dx=D\sqrt{\frac{\tilde{p}_0}{\delta}-\sqrt{\frac{\tilde{p}_{0}^2}{\delta^2}-\frac{\kappa^2}{\delta}}}cos(\sqrt{\delta}x),\nonumber\\
&M^y(x)=Dy=-\frac{\kappa D}{|\kappa|}\sqrt{\frac{\tilde{p}_0}{\delta}+\sqrt{\frac{\tilde{p}_{0}^2}{\delta^2}-\frac{\kappa^2}{\delta}}}sin(\sqrt{\delta}x),\nonumber\\
&M^z(x)=Dncos(\theta)=0,
\end{eqnarray}
where $\tilde{p}_0=p_0/(\alpha D^2)$, $\kappa=\frac{C}{\alpha D^2}$, and $\delta=\frac{g_0+g_2D^2}{\alpha D^2}>0$.
By Eq.(\ref{89}), the frequency and spatial period of magnetization  are given by
\begin{eqnarray}
&\omega=\sqrt{\delta},\nonumber\\
&T=2\pi/\omega=2\pi /\sqrt{ \delta},
\end{eqnarray}
It shows that the spatial periods of magnetization and density are determined by interaction parameter $\delta$.
As $\delta$ gets smaller, the periods get larger.
If $\delta <0$, the period would be imaginary, which indicates system's instability.

 From Eq. (\ref{89}), it is evident that the period of magnetization is twice that of the particle number density, $n(x)$. When the particle orbits are closed, the spatial periods of all three magnetization components are the same (or commensurate). In order to illustrate this, we have plotted a closed elliptical orbit in Fig. 2 with a current density of $J_0=0$. Additionally, in Fig. 3, we have plotted the functions of $x$, $y$, and $z$ as a function of time, $t$.

\begin{figure}[th]
%æ¯å¯éé¡¹ hè¡¨ç¤ºçæ¯hereå¨è¿éæå¥ï¼tè¡¨ç¤ºçæ¯å¨é¡µé¢çé¡¶é¨æå¥
\centering
\includegraphics[scale=0.45]{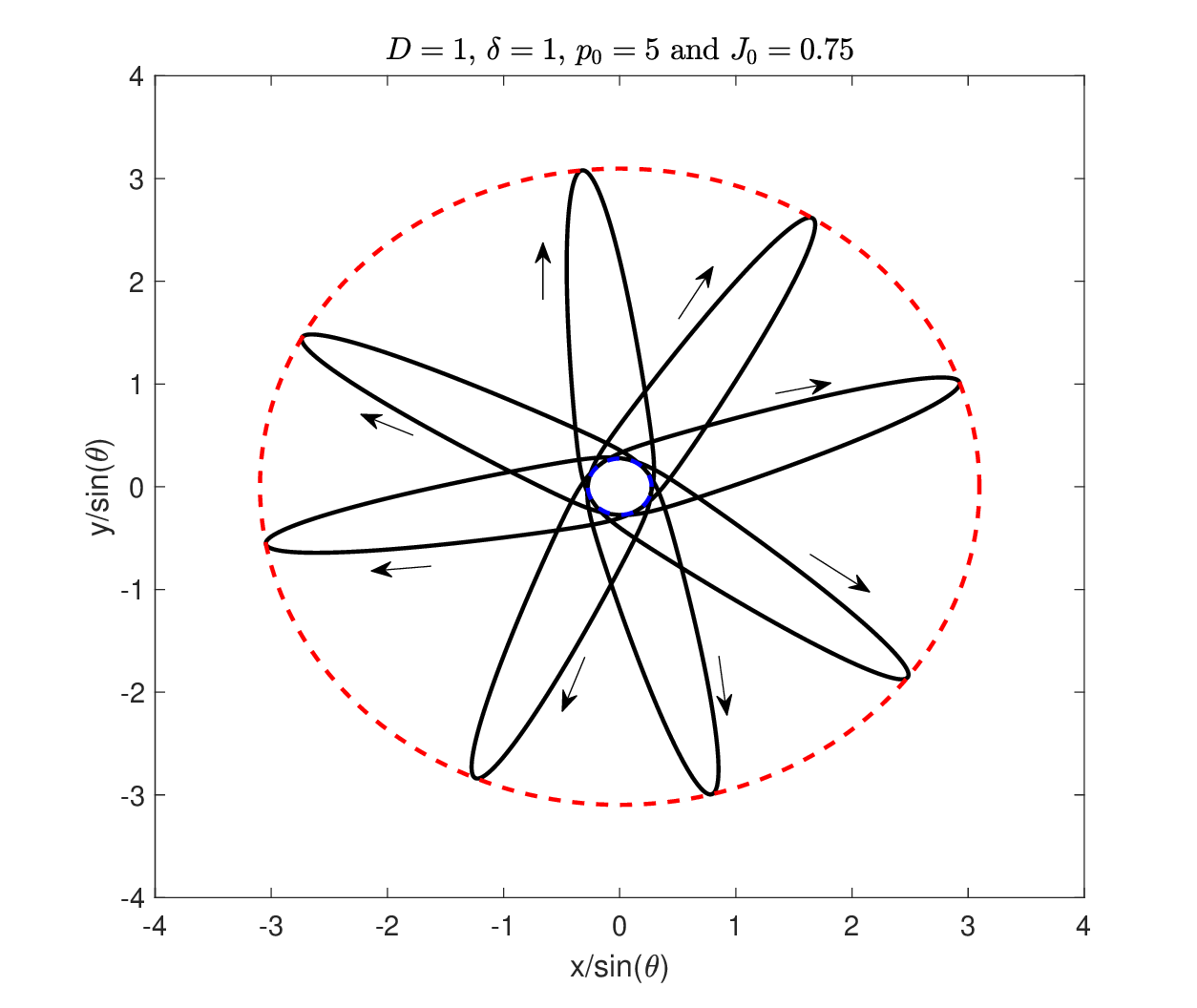}
\caption{The orbit of $x-y$ plane  with $p_0=5$ and $J_0=0.75$, $sin(\theta)=0.66$.  The arrows indicate the directions of the particle's motions.}
\label{fig:1}
\end{figure}

\begin{figure}[th]
%æ¯å¯éé¡¹ hè¡¨ç¤ºçæ¯hereå¨è¿éæå¥ï¼tè¡¨ç¤ºçæ¯å¨é¡µé¢çé¡¶é¨æå¥
\centering
\includegraphics[scale=0.45]{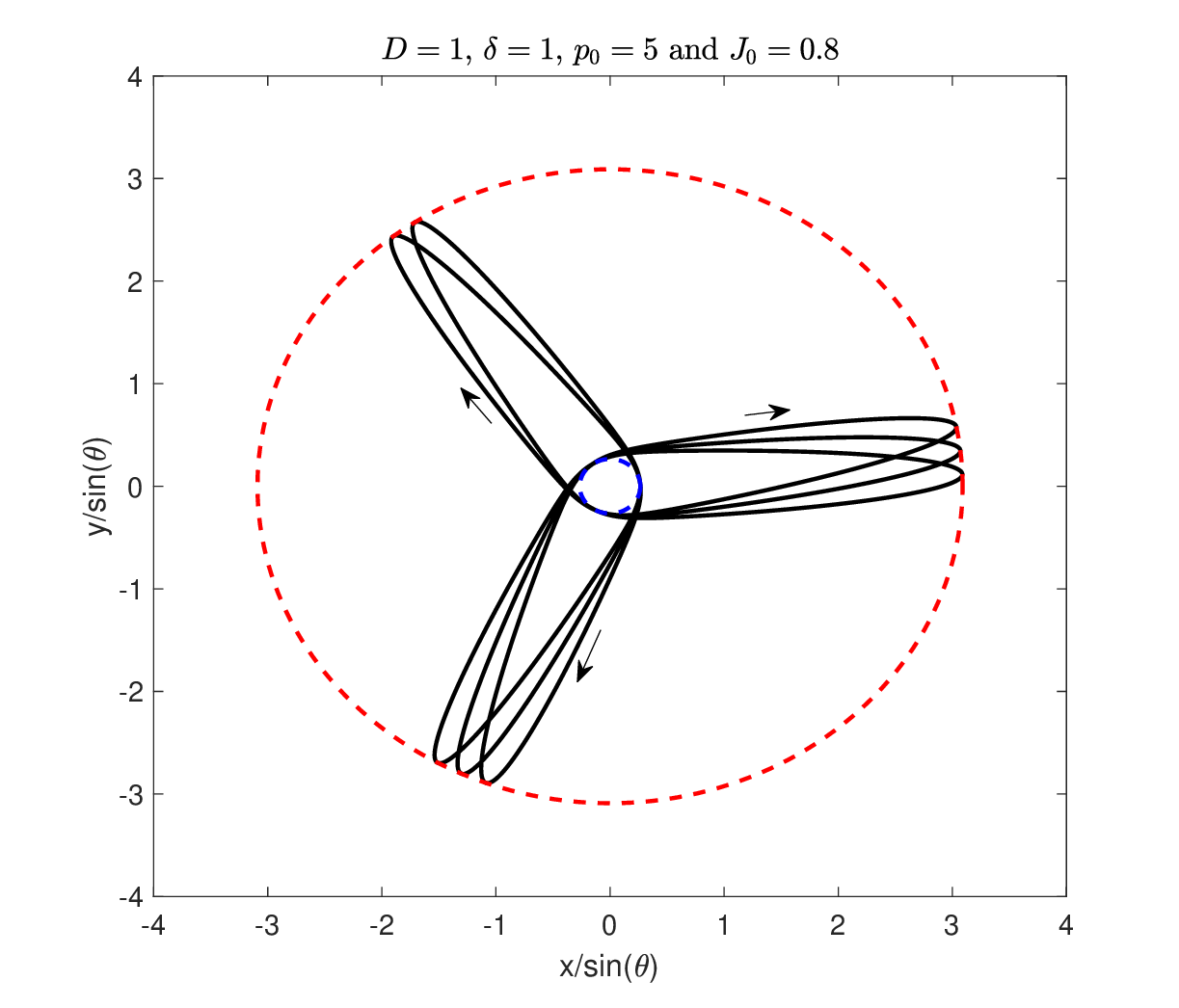}
\caption{The orbit of $x-y$ plane  with $p_0=5$ and $J_0=0.80$, $sin(\theta)=0.6$.  The arrows indicates the directions of the particle's motions.}
\label{fig:1}
\end{figure}

\subsection{$J_0\neq0$ (orbits are not closed) }
From Eq.(\ref{61}) and  Eq.(\ref{281}), we see that when $J_0\neq0$, then,   $\kappa= \frac{C}{\alpha D^2}$,  $\beta=\kappa^2sin^2(\theta)$, $cos(\theta)\equiv\frac{J_0 D}{C}\neq0$, $M^z=Dncos(\theta)\neq0$, $\gamma=\frac{J^{2}_{0}}{\alpha D^2}\neq0$.
% Due to the complex of this expression, we would not present here.
In such a case, the configuration space of particle's motion would be circular cone surface (polar angle $\theta=Const$) with a
vertex at the origin (force center).
It is important to note that the orbit would not be closed in general and would instead precess over time. To observe this precession in the x-y plane, we can use Eq.(\ref{94}) to define a precession angle, $\Delta \phi$, over two periods of distance $r(t)$, specifically $r_{min}\rightarrow r_{max}\rightarrow r_{min}\rightarrow r_{max}\rightarrow r_{min}$, i.e.,
\begin{eqnarray}
&\Delta \phi=2\pi+4\int_{r_{min}}^{r_{max}}\frac{-\kappa dr}{r^2\sqrt{2[\tilde{p}_0-(\frac{\beta}{2r^2}+\frac{\gamma}{r}+\frac{\delta }{2}r^2)]}}.
\end{eqnarray}
When $J_0=0$, $\gamma=0$, $\Delta \phi=0$,
so the precession angle is zero (see Fig.2).

%When $J_0\neq0$, $\gamma\neq0$, the orbit would have finite precession angle.

%we use units of system $m=\alpha=C^z=1$, and

In the following, we take some specific parameters $J_0=0.25,0.75,0.8,0.999$ to plot some figures (see Figs. 4-8).
From Eq.(\ref{2311}) and Eq.(\ref{J0}), we can see that when current density  $J_0$ varies from $0$ to $C/D$  , the polar angle $\theta$ varies  from $\pi/2$ to $0$ and $sin(\theta)$ goes from $1$ to $0$.
 Consequently,  $x,y-$ components of magnetization get smaller and smaller and $z-$ component magnetization would dominate [see Eq.(\ref{791})] .
 In order to see the precession of the orbit of  $x-y$ plane, we magnify the scales  of orbits of $x-y$ plane  by multiplying a factor $1/sin(\theta)$ in the figures of orbits (Figs.5-8).

Because the density $n(x)$ is a periodic function of $x$, we can observe from Eq.(\ref{791}) that the $z$-component of magnetization is also a periodic function of time with the same period as $r$. As shown in Fig.5, when the particle current density $J_0$ is small, the orbit in the $x-y$ plane (corresponding to the $x,y$-components of magnetization $M^{x,y}$) forms a Hooke's ellipse with a small precession angle. This implies that the spatial periods of $x,y,z$ components of magnetization are incommensurate (see Fig.4).

\begin{figure}[t]
%æ¯å¯éé¡¹ hè¡¨ç¤ºçæ¯hereå¨è¿éæå¥ï¼tè¡¨ç¤ºçæ¯å¨é¡µé¢çé¡¶é¨æå¥
\centering
\includegraphics[scale=0.45]{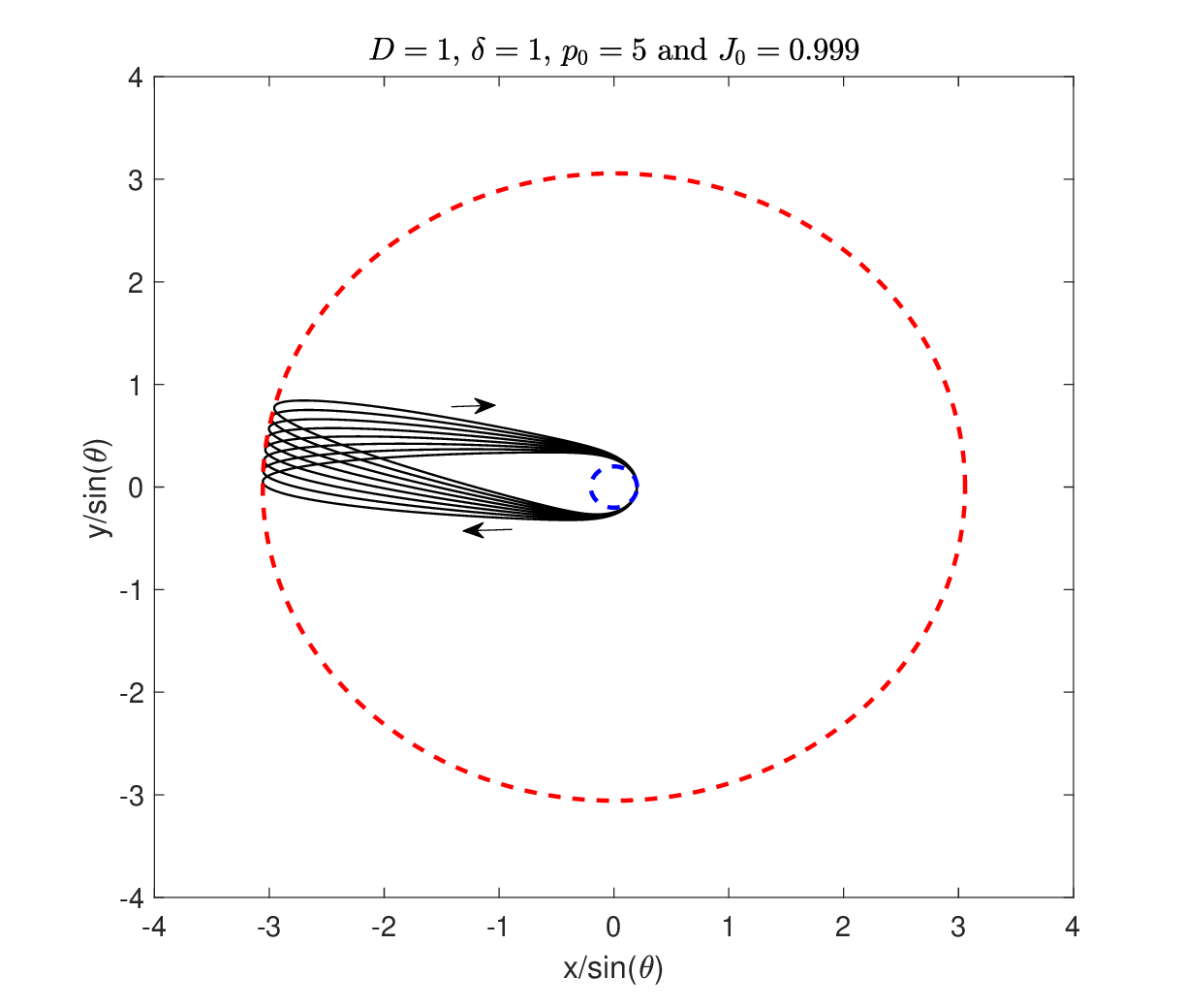}
\caption{The orbit of $x-y$ plane  with $p_0=5$ and $J_0=0.999$, $sin(\theta)=0.045$.  The arrows indicate the directions of the particle's motions. }
\label{fig:1}
\end{figure}

\begin{figure}[!h]
%æ¯å¯éé¡¹ hè¡¨ç¤ºçæ¯hereå¨è¿éæå¥ï¼tè¡¨ç¤ºçæ¯å¨é¡µé¢çé¡¶é¨æå¥
\centering
\includegraphics[scale=0.43]{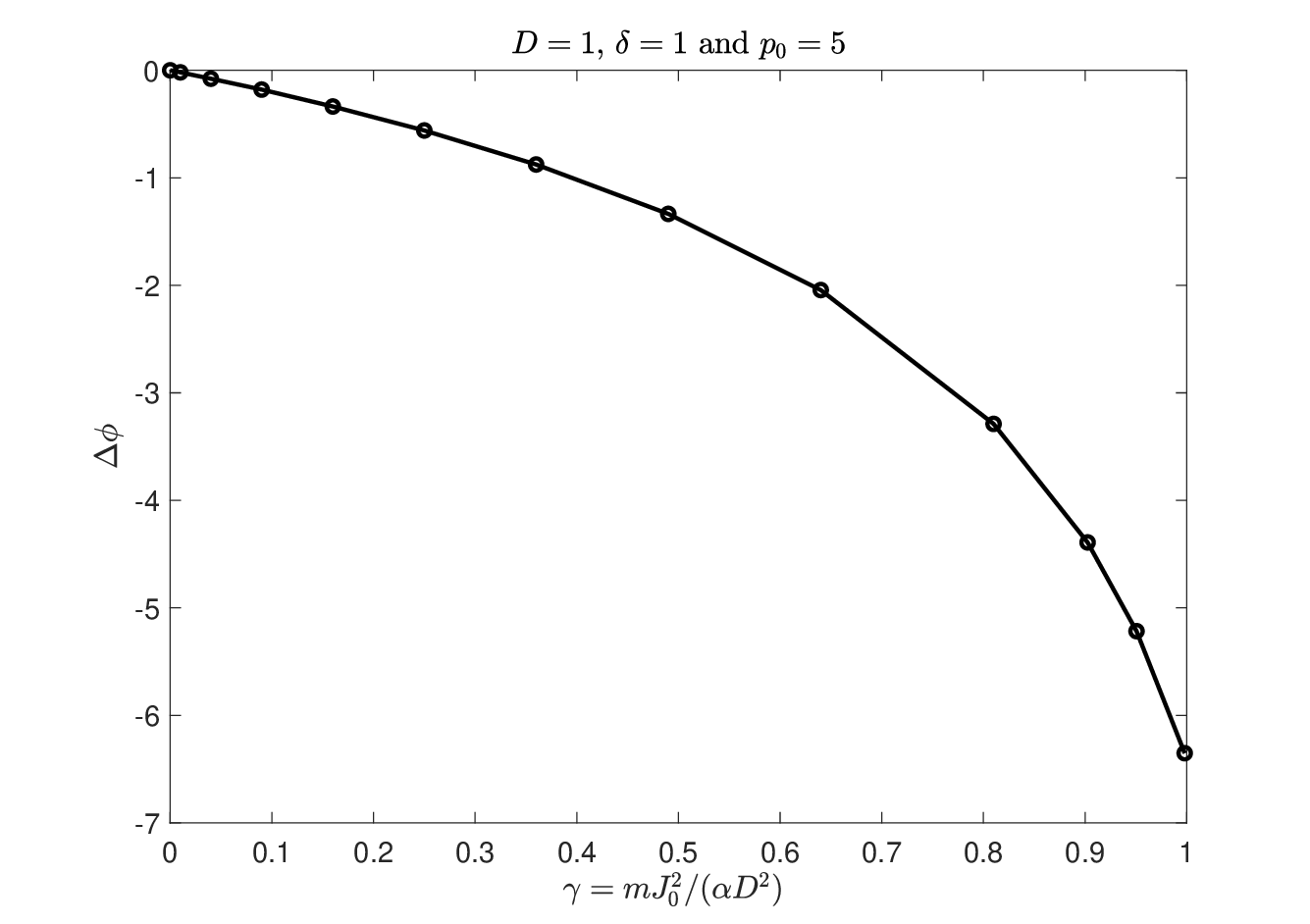}
\caption{The precession angle of the orbit as a function of Coulomb potential strength $\gamma =mJ_{0}^{2}/(\alpha D^2)$.  The negative sign of the angle indicates that the orbit precesses in a clockwise direction in the $x-y$ plane. When $\gamma$ is small, as the value of $\gamma$ increases, the precession angle also increases proportionally.}
\label{fig:1}
\end{figure}

Fig.9 illustrates the relationship between the precession angle and the strength of the Coulomb potential $\gamma$. It is evident that as the current $J_0$ ( or $\gamma$) increases, the precession angle also increases (see Fig.9 ).  Furthermore, we observe that for small values of $J_0$, the precession angle is directly proportional to the strength of the Coulomb potential $\gamma$ (see Fig.9), i.e.,
\begin{eqnarray}
&\Delta \phi\propto \gamma= mJ_{0}^{2}/(\alpha D^2).
\end{eqnarray}
These results show that the repulsive Coulomb  potential $\gamma/r$ of Eq.(\ref{80}) results in the precession of the orbits and  a mass superlow $J_0\neq0$ usually induce an incommensurate magnetization in a U(2) invariant superfluid.

\section{Conclusions}\label{sec:5}
In summary, we have investigated the one-dimensional steady solution to the hydrodynamic equation in a U(N) invariant superfluid at zero temperature. Our findings show that the magnitude of magnetization is always directly proportional to the particle number density. In the case of U(2), the one-dimensional steady solution can be mapped to a problem of a particle's motion in a central force field. The particle's distance from the origin is a periodic function of time. When the particle current is zero, the particle's orbit is closed, resulting in the spatial periods of the three magnetization components being the same and commensurate. However, when the particle current is nonzero, the orbit is generally not closed, and the periods of the three components of magnetization are usually not commensurate. This suggests that a mass superflow typically induces an incommensurate magnetization. These results have potential implications for cold atom experiments.
Additionally, there are other questions that warrant further investigation in U(N) superfluids. For instance, exploring how to extend these findings to a U(3) invariant superfluid would be an interesting and challenging endeavor.

\section{Acknowledgments}
This work was supported by the NSFC under Grants Nos.
11874127, the Joint Fund with
Guangzhou Municipality under No.
202201020137, and the Starting Research Fund from
Guangzhou University under Grant No.
RQ 2020083.

\section{References}

\end{document}